\documentclass{svproc}
\usepackage{graphicx}%
\usepackage{multirow}%
\usepackage{amsmath,amssymb,amsfonts}%
\usepackage{mathrsfs}%
\usepackage[title]{appendix}%
\usepackage{xcolor}%
\usepackage{textcomp}%
\usepackage{manyfoot}%
\usepackage{booktabs}%
\usepackage{algorithm}%
\usepackage{algorithmicx}%
\usepackage{algpseudocode}%
\usepackage{listings}%
\usepackage{url}

\def\orcidID#1{\unskip$^{[#1]}$} 
\newcommand{\specialcell}[2][c]{%
  \begin{tabular}[#1]{@{}c@{}}#2\end{tabular}}
\newcommand{\legendbox}[1]{%
  \textcolor[HTML]{#1}{\rule{\fontcharht\font`X}{\fontcharht\font`X}}%
}

\usepackage{hyperref}

\usepackage{wrapfig}

\begin{document}
\mainmatter              
\title{Decoding Decentralized Finance Transactions through Ego Network Motif Mining}
\titlerunning{Decoding DeFi Transactions through Ego Network Motif Mining}  
%
\author{Natkamon Tovanich\inst{1}\orcidID{0000-0001-9680-9282} \and \\
Célestin Coquidé\inst{2}\orcidID{0000-0001-8546-6587} \and
Rémy Cazabet\inst{2}\orcidID{0000-0002-9429-3865}}
\authorrunning{Tovanich, Coquidé, and Cazabet} 

\institute{CREST, CNRS, École Polytechnique, Institut Polytechnique de Paris,\\91120 Palaiseau, France
\email{natkamon.tovanich@polytechnique.edu}\\
\and
LIRIS, CNRS, Université Claude Bernard Lyon 1, 69622 Villeurbane, France
\email{celestin.coquide@liris.cnrs.fr}, \email{remy.cazabet@gmail.com}
}

\maketitle              

\begin{abstract}
Decentralized Finance (DeFi) is increasingly studied and adopted for its potential to provide accessible and transparent financial services. Analyzing how investors use DeFi is important for reaching a better understanding of their usage and for regulation purposes. However, analyzing DeFi transactions is challenging due to often incomplete or inaccurate labeled data.
This paper presents a method to extract ego network motifs from the token transfer network, capturing the transfer of tokens between users and smart contracts.
Our results demonstrate that smart contract methods performing specific DeFi operations can be efficiently identified by analyzing these motifs while providing insights into account activities.

\setcounter{footnote}{0} 

\end{abstract}
\section{Introduction}
\label{sec:intro}

Decentralized finance (DeFi) is a financial infrastructure implemented on smart contracts in blockchains \cite{auer2024technology}. DeFi replicates traditional financial services without centralized intermediaries.
The building blocks of DeFi applications consist of tokens, protocols, and aggregators.
Tokens are digital assets created on the blockchain that represent value or utility. Tokens can be transferred from any sender account (i.e., a user or a protocol) to a recipient account.
In DeFi protocols, tokens provide specific financial services, such as borrowing, lending, and exchanging tokens. The user creates a ``transaction'' by calling a ``method'' (later, DeFi method) in the protocol's smart contract to perform a specific DeFi operation. A single DeFi method can involve one or multiple token transfers.
These protocols can be ``composed'' to create more complex financial products and strategies \cite{kitzler2023disentangling}. For instance, aggregators are platforms that integrate multiple DeFi protocols, typically allowing their clients to manage complex operations automatically to optimize their performance, e.g., automatically choosing the most favorable exchange rate or lending rate among multiple platforms.



As DeFi applications keep growing and evolving rapidly, it becomes crucial to understand the nature and usage of its protocols and analyze users' activities.
To handle this, we investigated ego networks of token transfers sent from (or received by) a particular account.
This work addresses two research questions which, to the best of our knowledge, have not been analyzed in previous work.

\begin{enumerate}
    \item Can we infer DeFi methods (e.g., depositing, borrowing, swapping, etc.) from the ego token transfer network?
    \item Can we identify users with similar activities using ego network motifs?
\end{enumerate}

Analyzing DeFi methods that perform a specific operation within Ethereum transactions usually requires data in which the DeFi methods are labeled, obtained from Etherscan, private companies, or smart contract interfaces (Application Binary Interfaces, ABIs).
While our study utilizes these DeFi method labels obtained through self-reported or manual labeling, we recognize this as a limitation, particularly when dealing with transactions that have missing or inaccurate labels.
By leveraging ego network motifs, our approach aims not only to provide a more accurate and interpretable tool for inferring DeFi methods but also to analyze trading activities of accounts within DeFi ecosystems, even when labeled data is incomplete or unavailable.


\section{Related Work}
\label{sec:related}

The study of blockchain networks has gained significant interest due to the rise of cryptocurrencies and DeFi applications. Many works study macroscopic properties, such as degree distribution or clustering coefficient, in various network representations \cite{khan2022graph,wu2021analysis}. Research on meso and microscopic behaviors is emerging, focusing on illegal activities in Bitcoin (e.g., \cite{akcora2019bitcoinheist,bellei2024shape}) and Ethereum (e.g., \cite{chen2020survey,chen2020phishing,wu2023towards}).

Network motifs play a crucial role in characterizing different groups of behaviors within these networks.
Kosyfaki et al. proposed an algorithm to identify significant flow motifs in the large Bitcoin user network \cite{kosyfaki2018flow}, while Ba et al. defined triadic closure network motifs and explored their impact on the evolution mechanisms of platforms \cite{ba2023characterizing}.
Temporal motifs have been employed in various tasks, including detecting mixing services in Bitcoin \cite{wu2021detecting}, analyzing dark web marketplaces and NFT wash tradings \cite{arnold2024insights}, detecting phishing scam accounts \cite{wang2023detecting}, and identifying account communities \cite{ao2021temporal,tovanich2023fingerprinting}.
Building on these works, we introduce ego network motifs, demonstrating their utility in extracting DeFi method signatures and characterizing account behaviors.

In addition to behavioral analysis, our research contributes to understand DeFi trading strategies. Zhou et al. proposed two graph algorithms to identify cyclic arbitrage in DeFi tradings and calculate the optimal threshold of profitable transactions \cite{zhou2021just}. The majority of the work in this domain has focused on Maximal Extractable Value (MEV), which measures the maximum profit that can be extracted by reordering, including, or excluding transactions within a block \cite{daian2019flash,qin2022quantifying}.
EigenPhi is a data platform that tracks these transactions in real time and provides graphs to analyze profitability.\footnote{\url{https://eigenphi.io/}}
Park et al. developed a graph neural network model to detect MEV transactions from token transfer graphs without relying on smart contract code or ABIs \cite{park2024unraveling}.

Our work extends this body of research by proposing a novel motif extraction technique to infer DeFi methods. These motifs serve as building blocks for analyzing more complex interactions, including arbitrage and MEV transactions.

\section{Data Description}
\label{sec:data}

In this work, we focused on a subset of Ethereum accounts labeled on Etherscan as likely trading with decentralized finance (DeFi) protocols.
We obtained the list of all ``fund'' accounts from Etherscan\footnote{\url{https://etherscan.io/accounts/label/fund}} as of March 2023, which consists of 65 entities. Among these, 25 are identified as associated with Alameda Research, a cryptocurrency trading firm linked to FTX that was active in DeFi before the collapse of both entities in November 2022 due to massive fraud.
This subset was chosen to examine the DeFi interactions of notable trading accounts rather than to analyze the entire blockchain.
Although this list does not cover all accounts, it provides a focused perspective on significant players in the DeFi space.

\vspace{2mm}

\noindent\textbf{Token transfer dataset:} A transfer is an action of moving Ether (ETH) or tokens from one account to another. Each transaction can trigger one or multiple transfers. An account can be either an address controlled by a user account (Externally Owned Account, EOA) or a smart contract (CA).
The special address 0x00000..., known as NULL, is commonly used for minting and burning tokens.

\begin{figure}
\centering
\includegraphics[width=\textwidth]{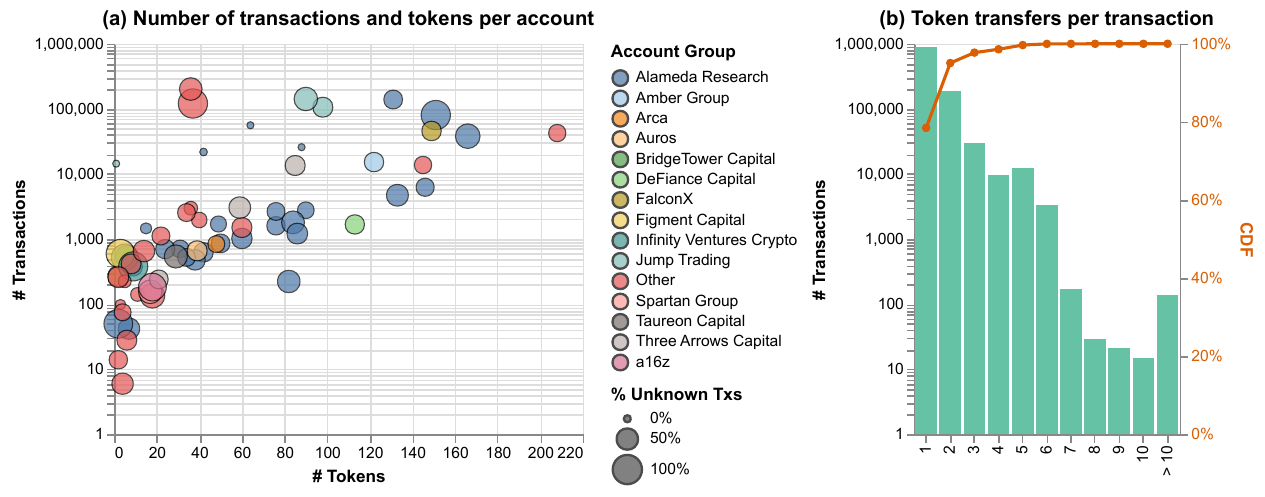}
\caption{(a) Fund accounts in our dataset and (b) Token transfers histogram}
\label{fig:accounts}
\vspace{-5mm}
\end{figure}

To construct the ego transfer network (ETN), token transfer data can be collected from Transfer API from Alchemy\footnote{\url{https://docs.alchemy.com/reference/transfers-api-quickstart}}. This API provided detailed records of all token transfers sent or received by our selected accounts, including transaction hashes, sending and receiving accounts, token names, and transfer amounts.
We collected token transfer data from the 65 selected accounts, resulting in 1,598,098 transfers across 1,131,723 transactions.
\autoref{fig:accounts} describes key elements of the dataset: \autoref{fig:accounts} (a) shows, for each account, its number of transactions, number of different tokens used, and fraction of transactions of unknown nature. \autoref{fig:accounts} (b) describe the number of token transfers per transaction. We observe that most transactions consist of a single token transfer (78\%), while more than a hundred thousand transactions have more than one single token transfer.


\vspace{2mm}

\noindent\textbf{Token list:} Each blockchain has its native cryptocurrency, such as Bitcoin (BTC) on Bitcoin, Ether (ETH) on Ethereum, and MATIC on Polygon, allowing users to transfer tokens within the network. Besides its native token, Ethereum blockchain allows the creation of other tokens by implementing smart contracts that follow standards like ERC-20 (fungible), ERC-721 (non-fungible or NFT), and ERC-1155 (multi-token).
Fungible tokens can be divided into varied-price cryptocurrencies (e.g., ETH, DODGE, SHIB) and stablecoins (e.g., USDC, USDT, DAI), which are pegged to a fiat currency like the US Dollar.
Synthetic tokens are created to represent underlying assets, often through mechanisms such as wrapping, staking, or collateralization within DeFi protocols.

We listed all tokens in our token transfer dataset and applied the following data processing steps to filter out spam tokens and assign categories:

\begin{enumerate}
    \item \textbf{Assigning token categories:} We manually verified the top 500 tokens in our dataset based on the number of transactions and assigned their category based on Messari's classification\footnote{List of tokens: \url{https://messari.io/assets?view=e88a221b-cb05-4995-96a7-cf2c5e217e79} Classification scheme: \url{https://docs.messari.io/docs/messari-classification-system}}.
    \item \textbf{Filtering out spam tokens:} Spam tokens are created to deceive the recipient into trading or interacting with valueless or harmful contracts. These spam tokens usually replicate the name and symbol of legitimate tokens or well-known brands.
    We checked if the token was on Etherscan’s list of verified tokens labeled\footnote{\url{https://etherscan.io/tokens}}. If a token was not in this list, we removed transactions where such a spam token was transferred from our analysis.
    \item \textbf{Unlabeled tokens:} Remaining tokens were annotated as ``Unlabeled''. Although their specific purpose is unknown, we included them in our analysis since they are not considered as spam tokens.
\end{enumerate}

\begin{table}[ht]
\caption{Token Categories}
\vspace{-2mm}
\label{tab:tokens}
\resizebox{\textwidth}{!}{
\setlength{\tabcolsep}{3pt}
\begin{tabular}{@{}l@{}r@{}r p{9cm}@{}}
\toprule
\multicolumn{1}{c}{\textbf{Token Category}} & \multicolumn{1}{c}{\textbf{\#Tokens (\%)}} & \multicolumn{1}{c}{\textbf{\#Txs (\%)}} & \multicolumn{1}{c}{\textbf{Top-5 Tokens (with \%Txs)}}\\
\midrule
Cryptocurrency    & \specialcell[t]{17  \\  (2.04\%)} & \specialcell[t]{483,038 \\ (35.94\%)} & ETH (17.70\%), WETH (14.43\%), WBTC (2.14\%), SHIB (1.40\%), PAXG (0.12\%)                      \\
Stablecoin        & \specialcell[t]{27  \\  (3.23\%)} & \specialcell[t]{385,452 \\ (28.68\%)} & USDT (15.03\%), USDC (8.93\%), BUSD (1.76\%), DAI (1.14\%), TUSD (0.46\%)                       \\
Marketplace       & \specialcell[t]{99  \\ (11.86\%)} & \specialcell[t]{223,954 \\ (16.66\%)} & COMP (1.50\%), UNI (1.46\%), SNX (1.19\%), CRV (1.04\%), CHI (1.01\%)                           \\
Other             & \specialcell[t]{100 \\ (11.98\%)} & \specialcell[t]{98,927  \\  (7.36\%)} & LINK (1.50\%), ENS (0.97\%), GRT (0.63\%), NMR (0.52\%), CVX (0.35\%)                           \\
NFT \& Metaverse & \specialcell[t]{57  \\  (6.83\%)} & \specialcell[t]{59,482  \\  (4.43\%)} & MANA (0.63\%), APE (0.58\%), CHZ (0.56\%), AXS (0.53\%), LOOKS (0.30\%)                         \\
Network           & \specialcell[t]{29  \\  (3.47\%)} & \specialcell[t]{35,737  \\  (2.66\%)} & MATIC (1.12\%), FTM (0.65\%), OMG (0.22\%), INJ (0.20\%), LRC (0.10\%)                          \\
Financial Service & \specialcell[t]{40  \\  (4.79\%)} & \specialcell[t]{19,794  \\  (1.47\%)} & CEL (0.32\%), BADGER (0.20\%), SXP (0.20\%), RAY (0.16\%), MTA (0.12\%)                         \\
Synthetic         & \specialcell[t]{339 \\ (40.60\%)} & \specialcell[t]{18,310  \\  (1.36\%)} & stETH (0.17\%), WNXM (0.13\%), UNI-V3-POS (0.07\%), XRPBULL (0.06\%), variableDebtUSDT (0.05\%) \\
Bridge            & \specialcell[t]{12  \\  (1.44\%)} & \specialcell[t]{16,279  \\  (1.21\%)} & REN (0.65\%), SYN (0.15\%), QNT (0.12\%), T (0.10\%), NU (0.08\%)                               \\
Unlabeled         & \specialcell[t]{115 \\ (13.77\%)} & \specialcell[t]{3,146   \\  (0.23\%)} & MIC (0.03\%), AKRO (0.03\%), XCN (0.01\%), POLY (0.01\%), EMB (0.01\%) \\
\bottomrule
\end{tabular}
}
\vspace{-4mm}
\end{table}

\autoref{tab:tokens} lists all token categories along with the number of tokens, transactions, and the top 5 tokens in each category. After excluding transactions involving spam tokens, we have 1,598,098 transfers across 1,095,374 transactions.

\vspace{2mm}

\noindent\textbf{Labeled transaction dataset:} So far, we know token transfers within the transaction but don't know the DeFi method. Thus, we collected the transaction headers from Etherscan's transaction list pages for each account\footnote{e.g., \url{https://etherscan.io/txs?a=0x0f4ee9631f4be0a63756515141281a3e2b293bbe}}. The details of any transaction include the transaction hash, DeFi method, block number, timestamp, source and target accounts, ETH amount, and transaction fees.


However, we were able to retrieve DeFi method labels for 55\% of transactions, due to technical constraints of the Etherscan web crawling (limiting access to the latest 100,000 transactions per account and excluding transactions not created by the selected accounts).


\vspace{2mm}

\noindent\textbf{DeFi method list:} The \emph{method} column in the labeled transaction dataset specifies the DeFi method. We constructed the ground truth dataset from DeFi methods occurring in at least 100 transactions, manually grouping each DeFi method based on its label. \autoref{tab:methods} presents the method groups and their full names. We merged the ``Exchange'' and ``Swap'' method groups as well as the ``Redeem'' and ``Withdraw'' groups since they perform the same function.

\begin{table}[ht]
\vspace{-3mm}
\caption{Method Groups}
\vspace{-2mm}
\label{tab:methods}
\resizebox{\textwidth}{!}{
\setlength{\tabcolsep}{3pt}
\begin{tabular}{@{}l@{}r p{11.5cm}@{}}
\toprule
\multicolumn{1}{c}{\textbf{Group}} & \multicolumn{1}{c}{\textbf{\# Txs}} & \multicolumn{1}{c}{\textbf{Full Method Names with $\geq$ 100 Txs}}\\
\midrule
Transfer      & 404,130 & Complete Transfer, Safe Transfer From, Transfer, Transfer From, Transfer Tokens \\
Swap	      & 112,387 & \textbf{Swap:} Any Swap Out Underlying, Batch Eth Out Swap Exact In, Exact Input Single, Simple Swap, Swap, Swap ETH For Exact Tokens, Swap Erc20, Swap Exact ETH For Tokens, Swap Exact Tokens For ETH, Swap Exact Tokens For Tokens, Uniswap V3Swap, \textbf{Exchange:} Exchange, Exchange underlying \\
Withdraw      & 3,619   & \textbf{Withdraw:} Remove Liquidity ETH With Permit, Remove liquidity one coin, Withdraw, Withdraw Erc20, \textbf{Redeem:} Redeem, Redeem Underlying \\
Deposit       & 3,325   & Add Liquidity, Add Liquidity ETH, Deposit, Deposit ETH, Deposit For \\
Claim Reward  &	2,881   & Claim, Claim Comp, Claim Reward, Claim Rewards, Claim Token, Get Reward \\
Borrow	      & 1,389   & Borrow \\
Repay         & 1,256   & Repay, Repay Borrow \\
Mint          & 1,189   & Mint, Mint many \\
Exit          & 276     & Exit \\
Burn          & 241     & Burn \\
Stake         & 215     & Stake \\
\bottomrule
\end{tabular}
}
\vspace{-3mm}
\end{table}

We chose only the 8 DeFi method groups with more than 1,000 transactions in our dataset, ensuring a sufficient sample size for our classification models.
The final dataset results in 530,170 transactions (33\% of transactions in the token transfer dataset) with labeled DeFi method groups. The missing method data highlights the limits of relying only on labeled data, as is common in the literature, and thus the interest of our approach, which proposes to automatically infer functions from the transaction itself, even without the DeFi method label.

\section{Ego Network Motifs}
\label{sec:motif}


For each transaction, we construct the ego transfer network (ETN), represented as a directed heterogeneous graph.
Each account is represented as a node with an attribute denoting its account type. The account types can be Ego (E), address (A), contract (C), or the NULL address (N).
Each edge represents the token transfer directed from one account to another, with the edge attribute indicating the token category (see \autoref{tab:tokens}). We extract two features for each ego network: motif counts (denoted as $M$) and edge lists ($E$).

\vspace{2mm}

\begin{figure}
\includegraphics[width=\textwidth]{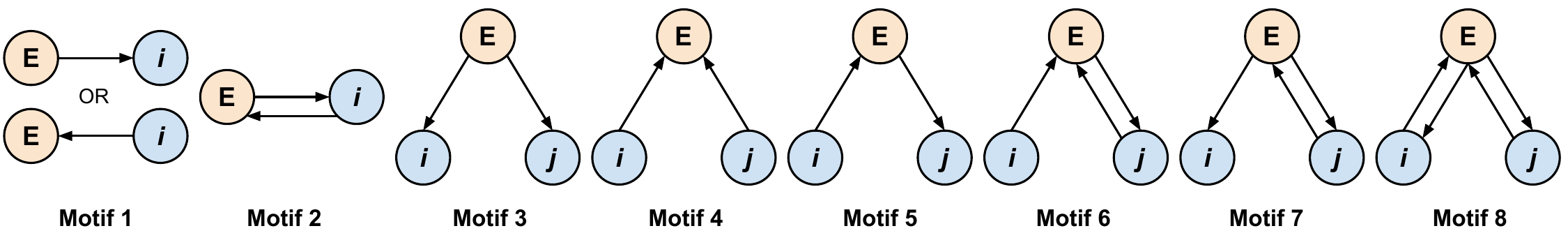}
\caption{List of all possible ego network motifs}
\label{fig:motifs}
\vspace{-4mm}
\end{figure}

\noindent\textbf{Motif counts:} Ego motifs from the ETN are directed subgraphs consisting of 2 or 3 nodes, represented in \autoref{fig:motifs}.
Note that, as our dataset records ego transfers---thus without edges between non-ego nodes---there are only 8 possible directed ego network motifs (out of 15 motifs in full networks).
The top node is always an ego node (E) connected to other nodes (denoted $i,j$).

We use the VF2++ algorithm \cite{juttner2018vf2} in NetworkX \cite{hagberg2008exploring}, to perform subgraph isomorphism matching to identify and count each occurrence of the motifs within the ETN.
We then check the node type for each motif found and represent it as a string with the pattern $\{motifID\}(E, i, j)$, where $\{motifID\} \in \{m1, m2, ..., m8\} $, with motif IDs corresponding to those in \autoref{fig:motifs}, while $i, j \in \{A, C, N\}$.

\vspace{2mm}

\begin{figure}
\includegraphics[width=\textwidth]{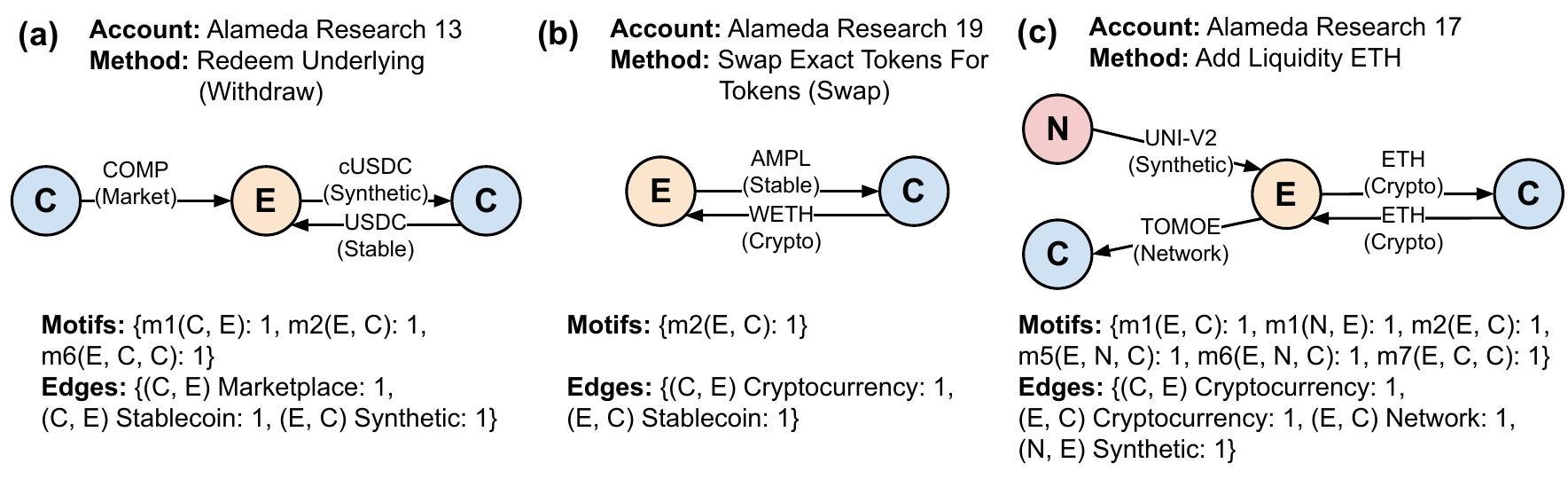}
\caption{Examples of transaction description by motifs and edge lists}
\label{fig:motif_examples}
\vspace{-5mm}
\end{figure}

\noindent\textbf{Edge lists:} Edge list indicates which token an ego received/spent from/to which account type. We obtain the edge list from the ego network with the token category. The edge count is represented as a vector $(S, T) \{token\;category\}$, where $S$ is the source node type and $T$ is the target node type.
Three examples of ETNs are presented in \autoref{fig:motif_examples}, along with motif counts and edge list vectors.

\section{DeFi Method Classification}
\label{sec:classification}

We use supervised learning methods to test whether motif counting and network features can infer DeFi methods (e.g., deposit, borrow, swap).
To determine which motif features can distinguish different DeFi methods, we tested four sets of training features: ego motifs with node types ($M$), edge list with token types ($E$), the combination of ego motifs and edge list ($M+E$), and the concatenation between motifs and edge list features within each motif ($M \times E$).

We trained three classification models: logistic regression (LR), decision tree (DT), and random forest (RF). For the LR model, we perform one-vs-all multiclass classifiers. To prevent overfitting while capturing patterns in the minority class, we set the minimum leaf node to 10 samples in both DT and RF. Given the dataset's high skew towards the most common DeFi methods (e.g., transfers and swaps in \autoref{tab:methods}), we assigned inversely proportional weights to the minority methods in the classification models to mitigate the class imbalance problem.

\begin{figure}
\vspace{-4mm}
\includegraphics[width=\textwidth]{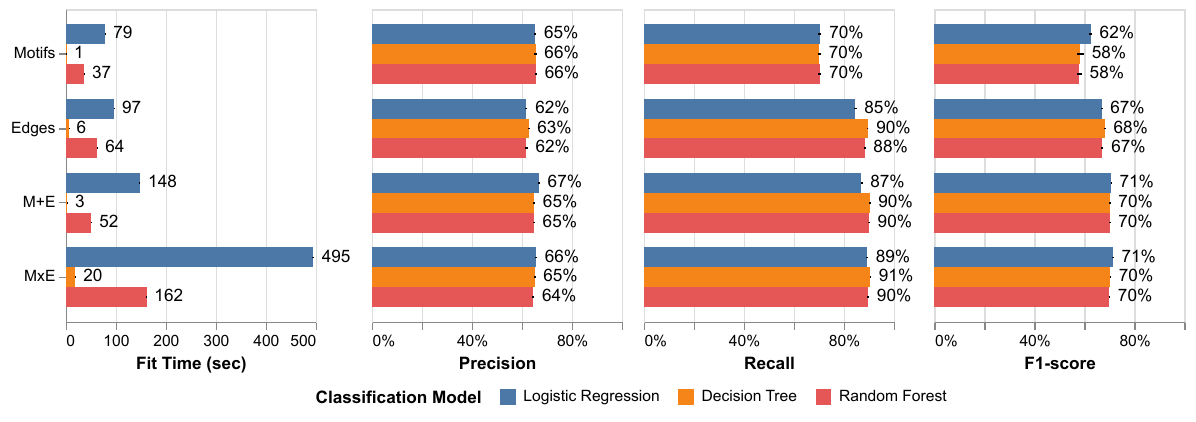}
\caption{Comparison of classification results with 10-fold CV scores}
\label{fig:classification}
\vspace{-5mm}
\end{figure}

We evaluated the classification models using stratified 10-fold cross-validation (10-fold CV), ensuring that each fold contained the same distribution of classes as the training set. The average F1-score, precision, and recall are reported in \autoref{fig:classification}, as accuracy is not a reliable metric for highly skewed classes.
The results indicate that using motifs with node and edge types can infer DeFi methods with moderate precision and high recall and outperform models using only motifs or edges.
The best model is LR with $M + E$ and $M \times E$ features, achieving an F-score of 71\%, followed by the DT and RF model with 70\%.
The $M \times E$ models did not significantly improve the F1-score compared to $M + E$ models, despite having a larger number of features.

\begin{wrapfigure}{r}{0.5\textwidth}
\includegraphics[width=\linewidth]{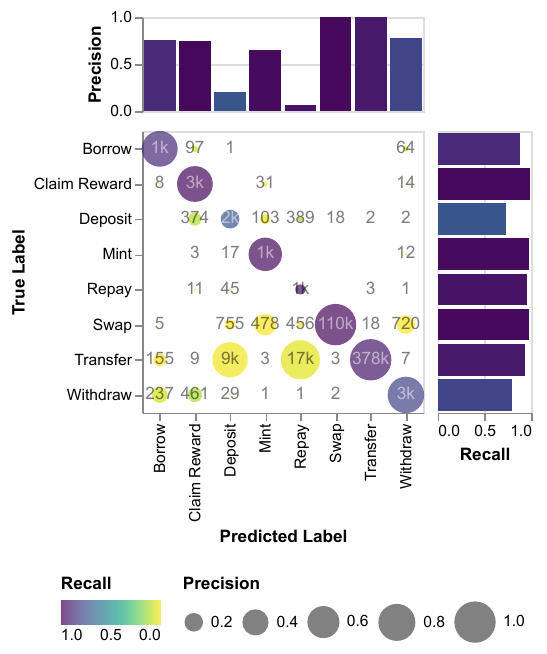} 
\caption{Confusion metric from the decision tree model using $M+E$ features}
\label{fig:confusion}
\vspace{-7mm}
\end{wrapfigure}

We chose to investigate the DT model with $M + E$ features further because it provides a simpler and more interpretable model while its performance is very close to that of the best-performing model.
Focusing on $M + E$ models, LR has slightly higher precision and F1-Score, while DT and RF had 3\% higher recall.

The confusion matrix in \autoref{fig:confusion} shows the number of samples predicted compared to the actual DeFi method. The color scale indicates the percentage of the true method for each predicted label (row-wise), while the circle size is proportional to the percentage predicted compared to the true class (column-wise). DT model achieves high recall for most DeFi methods, although it tends to mispredict transfer as deposit and repay, and swap as mint and withdraw, indicating the similarity in motif features and resulting in lower precision for these methods.

The classification results confirm that ego network motifs with account and token types are important features for inferring DeFi methods.

\section{Interpretable Method Signatures}


While the previous section has shown that one can use machine learning to classify DeFi methods from ego network motifs, the models remain a black box. In this section, we study which specific motif can be associated with which DeFi method to better understand the nature of tradings and make decisions more interpretable.
We used the decision tree (DT) model as a reference because it offered results comparable to the best-performing logistic regression (LR) model while achieving higher recall and maintaining interpretability. However, the full DT model is complex, making it challenging to interpret all the rules.

\begin{figure}
\centering
\includegraphics[width=\textwidth]{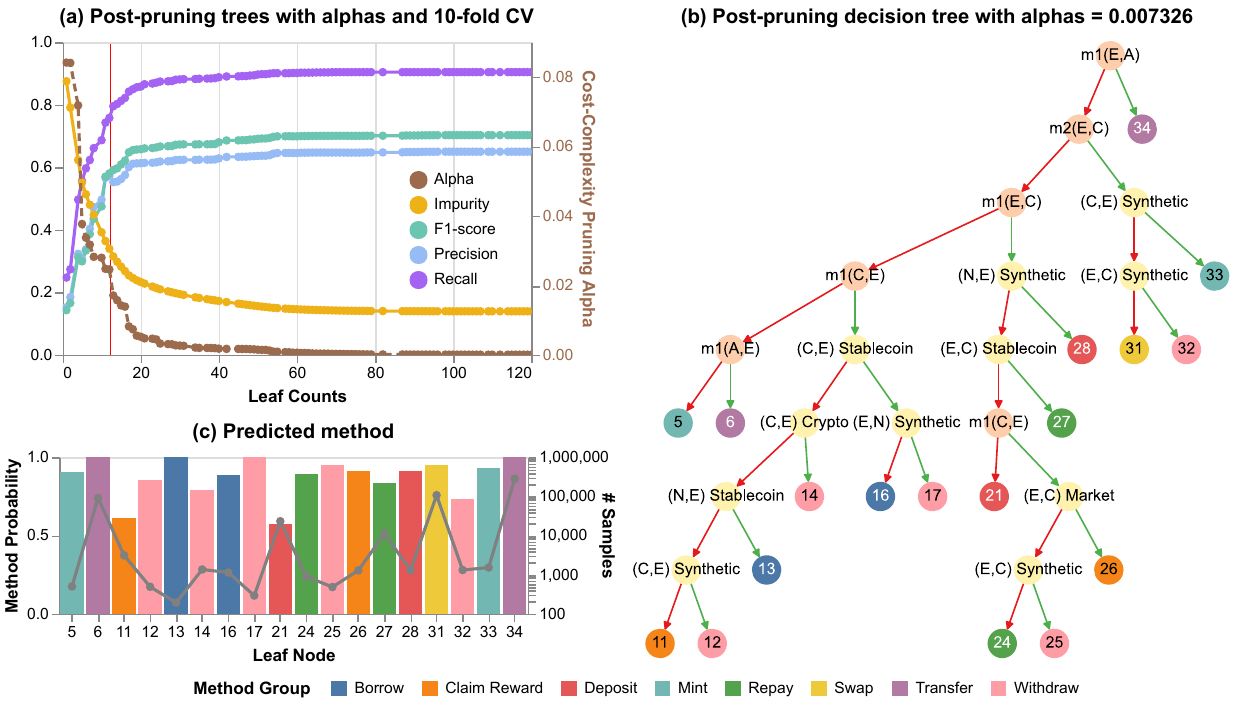}
\caption{(a) Post-pruning decision tree with varying alphas, leaf nodes, and 10-fold cross-validation metrics;
(b) Pruned tree selected with alpha = 0.007326;
(c) Predicted DeFi method at each leaf node (color), along with probability (bars) and sample size (line).}
\label{fig:post_pruning}
\vspace{-5mm}
\end{figure}

We performed post-pruning to reduce the tree's complexity and obtain a manageable set of rules while retaining predictive performance, thus simplifying the model. \autoref{fig:post_pruning} (a) shows post-pruning decision trees with varying cost-complexity pruning alphas and 10-fold CV scores for each number of leaf nodes. We chose a pruned tree with 18 leaf nodes (alpha = 0.007326), depicted in \autoref{fig:post_pruning} (b), while (c) shows the predicted class for each leaf node, with probabilities represented by bars and the number of samples (on a log scale) indicated by a gray line.

We adopt a frequent pattern mining approach to find a maximal set of motif features with a support of more than 80\%, i.e., features that appear in over 80\% of the instances at that leaf node.
For each leaf node, we start by listing motif features with support greater than the threshold, incrementally adding more features until no additional features meet the support criteria.
We selected the longest frequent itemset as the signature motif of the leaf and used it to analyze token transfer networks of unidentified transactions.

\begin{figure}
\vspace{-3mm}
\includegraphics[width=\textwidth]{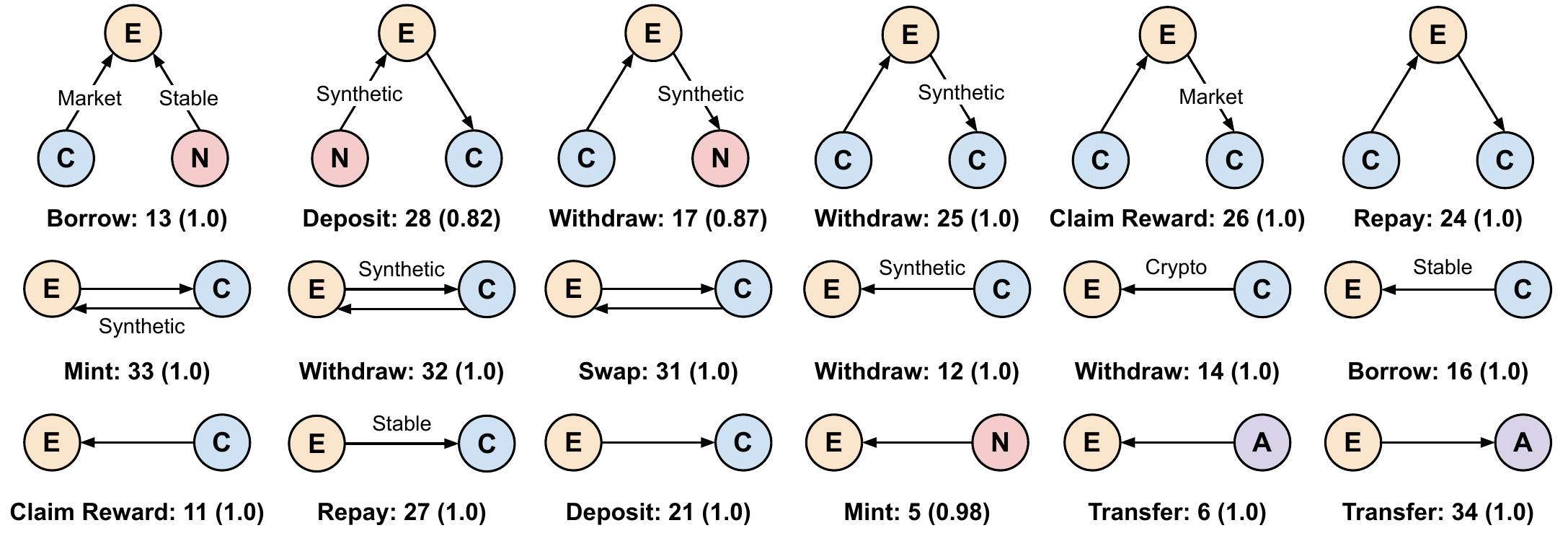}
\caption{List of DeFi method signature motifs on the leaf node with the predicted method. The support value for each motif is provided in parentheses.\\
\textit{Note: Crypto: Cryotocurrency, Stable: Stablecoin, Market: Marketplace tokens.}}
\label{fig:method_signatures}
\vspace{-5mm}
\end{figure}

\autoref{fig:method_signatures} displays the DeFi method signatures in 18 leaves of the pruned tree.
Most of the signature motifs reflect the functionality of the labeled methods.
The deposit (28) pattern indicates that an ego deposits tokens into a contract and receives synthetic tokens as collateral. Conversely, the withdraw method (17, 25) patterns show that an ego must burn or return synthetic tokens to withdraw their deposited tokens from the contract.
The swap method (31) pattern shows the exchange of tokens between the protocol's contract and the account. However, if the user receives (or sends) synthetic tokens to the contract, it corresponds to the mint (33) or withdraw (32) methods, respectively.

These signature motifs are interpretable, offering insights into transaction mechanisms while also being useful for investigating transactions with unknown DeFi methods.
However, we found that some signature motifs are overly specific to particular token categories, even though they could potentially apply to other tokens as well. For instance, the borrow (13, 16) patterns suggest that accounts (egos) in this dataset mostly borrowed stablecoins.
We anticipate that these patterns will become more precise and generalizable when we filter for known contracts associated with DeFi protocols or when using a larger dataset.

\section{Profiling Ethereum Accounts}
\label{sec:profiling}

\begin{figure}
\vspace{-3mm}
\includegraphics[width=\textwidth]{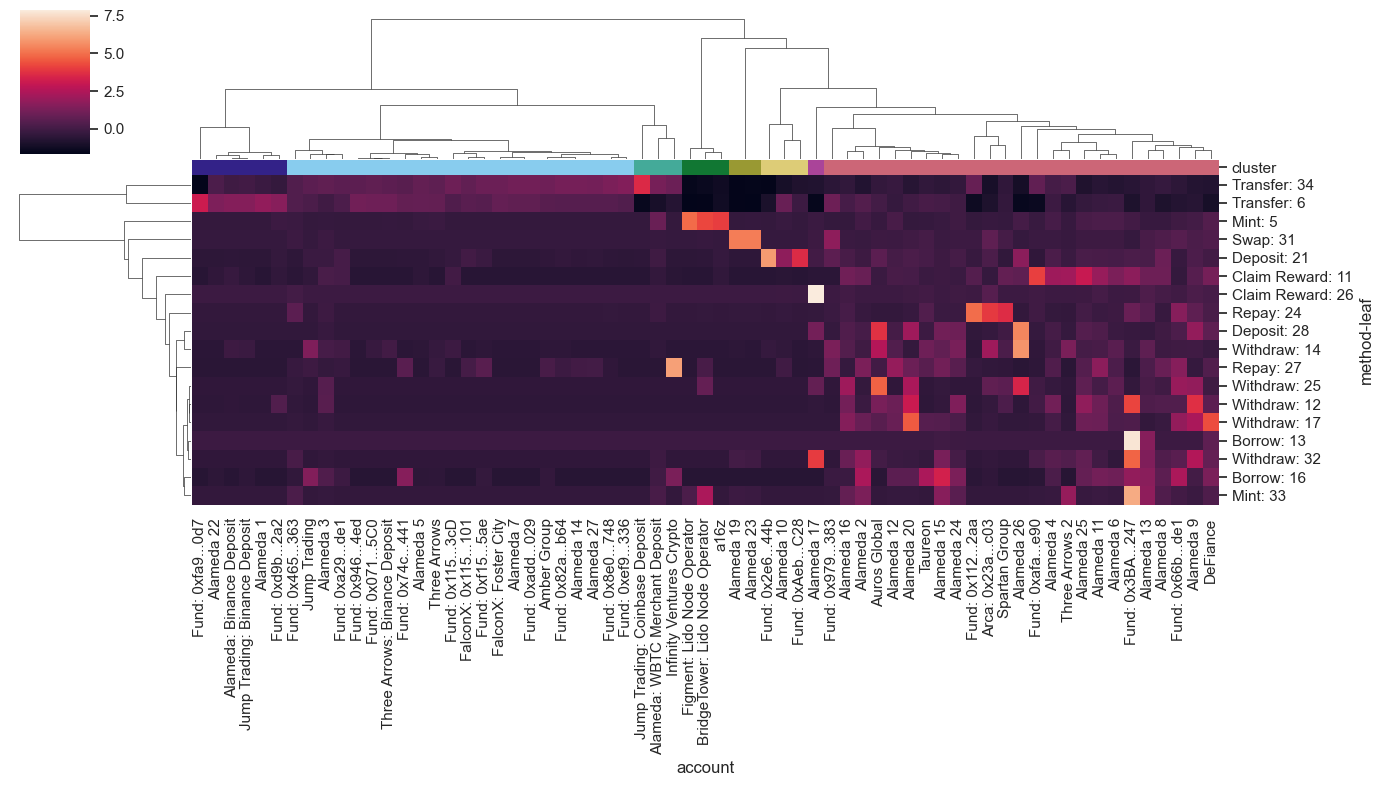}
\caption{Heat map and hierarchical clustering on fund account profiles} 
\label{fig:clustermap}
\vspace{-5mm}
\end{figure}

We demonstrate that extracted signature motifs can also be utilized to analyze DeFi activities at the account level. For each account, we create a profile by first counting the transactions associated with each DeFi method signature and then normalizing these counts by account.
\autoref{fig:clustermap} shows a hierarchical cluster map of accounts and method-leaf. Hierarchical clustering helps us detect groups of accounts with similar patterns of activity. Since some DeFi methods, like transfers, are overrepresented in all accounts, we show z-score normalization.

We use the silhouette score to determine the number of clusters (k = 8, silhouette = 0.42).
These 8 clusters are represented by different colors in the first row of the heat map in \autoref{fig:clustermap}, from Cluster 1 to Cluster 8 (left to right).

Cluster 1~\legendbox{332288} has a higher average z-score for transfer methods, typically receiving tokens from addresses (leaf 6), while Cluster 2~\legendbox{88CCEE} shows activity in both receiving and spending tokens (leaves 6 and 34). These accounts likely trading with off-chain exchange companies.
Cluster 8~\legendbox{AA4499} includes accounts primarily interacting with DeFi protocols, as indicated by a higher z-score for various methods, such as claim reward, withdraw, borrow, and repay.
Smaller clusters, with fewer than five accounts, are considered specialized. For instance, Cluster 5~\legendbox{999933} is characterized by higher token swap activities, suggesting interaction with decentralized exchanges.
Cluster 4~\legendbox{117733} is involved in token minting, while Cluster 6~\legendbox{DDCC77} can be identified as liquidity providers, focusing on deposit activities.

This section shows how motif signatures can be used to characterize accounts in the ego transfer network (ETN), despite unlabeled DeFi methods. These insights help us better understand the different groups of activities in our dataset.

\section{Conclusion and Future Work}
\label{sec:conclusion}


We demonstrated the effectiveness of using ego network motifs to represent and decompose token transfers in Ethereum transactions. By leveraging these motifs as fundamental building blocks, we successfully inferred DeFi methods and extracted method signatures to investigate transactions lacking labeled data. These motifs provide a powerful tool for analyzing account activities, allowing us to characterize different groups of accounts based on their DeFi activities.

The implications of our findings extend beyond academic research. For companies operating in the DeFi space, this approach offers a robust method for monitoring account activities and detecting trading strategies and anomaly behaviors.
For future work, we plan to expand this approach to analyze account activities over time, aiming to understand their responses to market fluctuations and shock events. Furthermore, we intend to apply our method to break down complex transactions, such as flash loans and arbitrage, and to incorporate dynamics and higher-order network motifs. This line of work will deepen our understanding of trading strategies and the evolution of the DeFi ecosystem, helping to identify emerging use cases without the need for fully labeled data.

\subsubsection*{Acknowledgement}
This work was supported by the CHIST-ERA grant FairOnChain, by Agence Nationale de la Recherche (ANR-23-CHRO-0002).
We thank Stefan Kitzler for his constructive feedback on this work.

%
%
\bibliographystyle{spmpsci} 
\bibliography{refs} 

\end{document}